# A Compact Permanent Magnet for Microflow NMR Relaxometry

Dmytro Polishchuk[*], Han Gardeniers[†]

University of Twente, 7500 AE Enschede, the Netherlands

We design and demonstrate a compact, robust and simple to assemble and tune permanent magnet suitable for NMR relaxometry measurements of microfluidic flows. Soft-magnetic stainless steel plates, incorporated inside the magnet airgap, are key for obtaining substantially improved and tunable field homogeneity. The design is scalable for different NMR probe sizes with the region of suitable field homogeneity, less than 200 ppm, achievable in a capillary length of about 50% of the total magnet length. The built physical prototype, having 3.5x3.5x8.0 cm$^3$ in size and 5 mm high airgap, provides a field strength of 0.5 T and sufficient field homogeneity for NMR relaxometry measurements in capillaries up to 1.6 mm i.d. and 20 mm long. The magnet was used for test flow rate measurements in a wide range, from 0.001 ml/min to 20 ml/min.

## 1. Introduction

Nuclear magnetic resonance (NMR), by far the most informative and versatile analytical technique used in chemistry, (bio-)medicine, food, and environment, is currently meeting drastic miniaturization [1–4] thanks to the advances in CMOS electronics [5–7] and (micro-)probe fabrication [8,9]. An emerging concept of miniature NMR sensors – NMR-on-a-chip [10–12] – is giving a promise of on-demand and on-line NMR applications, while substantially reducing the size, weight and cost of the equipment and hence making it portable and accessible.

One of the key components of NMR sensors is the source of required static magnetic field – magnets – which is required to be compact, lightweight, and providing sufficient field strength and homogeneity [13]. Superconducting magnets (SCM), commonly used in NMR studies and demanding cryogenic temperatures for operation, are bulky, energy consuming and too complex for portable NMR applications. Permanent magnets, on the other hand, are power independent and operating at room temperature, however, providing relatively weak fields, ranging from 0.5 to 1.5 T (compared with up to 28 T by SCMs [14]), and thus giving lower spectral resolution than SCMs. Recently demonstrated very compact permanent magnet designs [15,16] – of the size of a teacup and of sub-kg weight – can provide magnetic fields sufficient for NMR (bio-)chemical analysis of simple substances. The main problem of designing a suitable permanent magnet is achieving sufficient field homogeneity, which results in rather complex designs and additional shimming elements. Often, the assembly and mechanical tuning of such magnets is a rather demanding task [17,18] because of the strong sensitivity to minor displacements, non-uniform magnetization, material imperfections and fragility. Therefore, the pursuit for a suitable permanent magnet design as compact, efficient, simple in assembly and tuning as possible and with little or no electrical shimming is still ongoing.

In this work, we aimed at a compact permanent magnet system specifically designed for micro-fluidic flow measurements based on the NMR relaxometry principle. NMR relaxometry measures relaxation times of the nuclear spin system after excitation with an rf pulse – longitudinal, $T_1$, or transverse, $T_2$, with respect to the direction of the static magnetic field. Special spin-echo sequences, such as classic Carr-Purcell-Meiboom-Gill (CPMG) sequence [19,20], enable NMR relaxometry in relatively inhomogeneous magnetic fields – hundreds of parts per million (ppm) compared to <0.1 ppm required for low-resolution NMR spectroscopy. Therefore, our target field homogeneity was <200 ppm over a cylindrical volume of 1 mm i.d. and 20 mm long, whereas keeping the magnet length less than 10 cm (sensing volume length 25-50% of the total magnet length). Design and fabrication of such magnets, hardly possible a few decades ago, are facilitated nowadays by a number of modern materials and technologies, which we employ here: superior permanent magnet materials [21], such as families of Nd-Fe-B and Sm-Co materials giving strongest magnetic fields and good thermal stability, improved CNC (Computer Numerical Control) manufacturing machines with tolerances down to 10 μm (or better) and, finally, fast and reliable FEM (Finite-Element Method) numerical simulations.

---

[*] Also with Laboratory of Biophysics, Wageningen University & Research, 6708WE Wageningen, The Netherlands, and the Institute of Magnetism, NAS of Ukraine and MES of Ukraine, 03142 Kyiv, Ukraine.
[†] Corresponding author.
E-mail address: j.g.e.gardeniers@utwente.nl (H. Gardeniers).

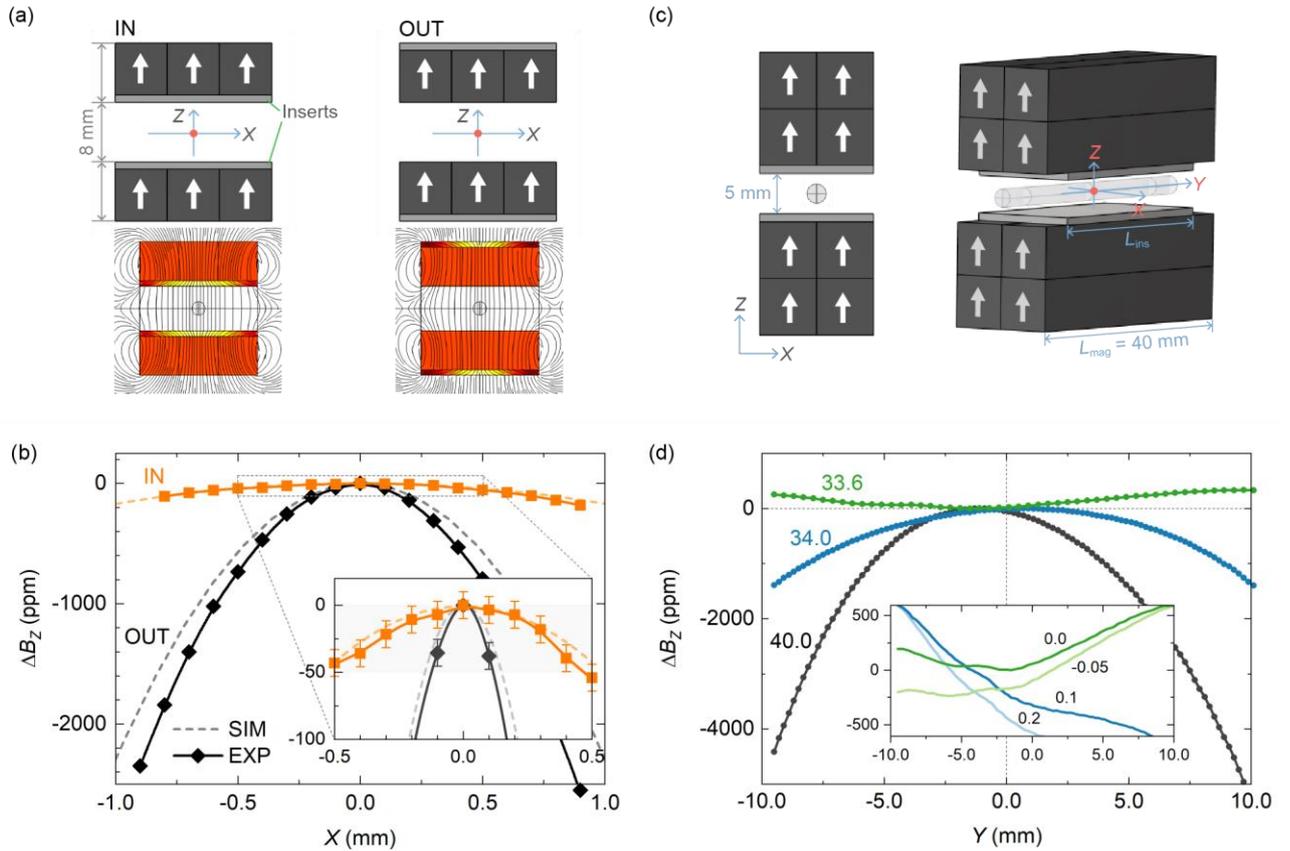

**Fig. 1.** Effect of soft-magnetic (stainless steel 410) inserts on field homogeneity in the airgap between two permanent magnets. (a) Two experimental settings when the 1-mm-thick inserts are inside (IN) or outside (OUT) the airgap and simulated distributions of magnetic field lines in transverse plane XZ with respect to the long side of the permanent magnet bars ($7 \times 7 \times 40$ mm$^3$). (b) Measured and simulated change in $Z$-component of magnetic flux density, $\Delta B_Z$, in lateral direction (axis X) in the airgap center ($Y, Z = 0$). (c) Experimental setting for studying $B_Z$ homogeneity in the longitudinal direction (axis Y). (d) Corresponding longitudinal $\Delta B_Z$ in the center of the airgap ($X, Z = 0$) for different insert lengths, $L_{ins}$ = 40, 34.0, and 33.6 mm or 1.0, 0.85, and 0.84$L_{mag}$, where $L_{mag}$ = 40 mm is the magnet length. Insert shows the change in $\Delta B_Z$-vs-$Y$ ($L_{ins}$ = 33.6 mm) when adjusting the insert off-center displacement along Y (values in mm).

Here we design and fabricate a compact ($3.5 \times 3.5 \times 8.0$ cm$^3$) and simple in assembly permanent magnet which demonstrates sufficiently strong (0.5 T) and uniform (<200 ppm) magnetic field suitable for microfluidic flow rate measurements in the sensing volume up to 40 μL (1.2 mm i.d. × 20.0 mm). Using this magnet, we measure actual flow rates in a range from 20 ml/min down to 0.001 ml/min, which can be of interest for micro- and nano-fluidic applications.

## 2. Magnet Design: Approach, prior experiments, and optimized prototype

***Approach and Prior Experiments.*** – Our approach is based on the fact that pieces of soft magnetic material (e.g. magnetic stainless steel, as used in this study) placed inside the airgap and adjoint to the two permanent magnets lead to a much more uniform magnetic field [22,23]. To obtain actual quantitative characteristics, a simple experiment was designed, as depicted in Fig. 1a. It compares magnetic field ($Z$-component of magnetic flux density, $B_z$, measured with a transverse Hall probe) in the center of the airgap along lateral axis X in between the two permanent magnets (each made of three $7 \times 7 \times 40$ mm$^3$ SmCo bar magnets) for two cases – when the soft-magnetic inserts (stainless steel, grade 410) are (i) inside or (ii) outside of the bore. As seen in Fig. 1b, when inside, the 1-mm-thick steel inserts improve the field homogeneity in the volume of interest (2 mm in diameter) by around 15 times, whereas the field strength somewhat decreases by around 35% (from 425 mT to 280 mT in the center). It is noteworthy that these experimental results are in remarkable agreement with the simulated results obtained for the same geometry and tabular material parameters using the COMSOL Multiphysics® software [24]. This assured us in reliability of our computer 3D model for the magnet design, which was confirmed later for other even more complex geometries.

The simulations also give an explanation to the observed substantially improved field homogeneity: owing to the soft magnetic properties, the steel inserts effectively redistribute the magnetic flux in the airgap's center (see inset in Fig. 1a). This distribution of flux lines is determined by the magnets' shape, induced magnetization direction and overall system geometry. Since this problem is complex to solve analytically, the FEM modeling, based on solving Maxwell's equation for the magnetic vector potential at each point [25], is indispensable and accurate enough for performing numerical experiments with different magnet designs.

Another experiment was designed for studying the effect of the steel inserts on the field homogeneity in the longitudinal direction – parallel to the long side of the bar magnets, axis Y in Fig. 1c. In order to make $B_z$ as uniform as possible along the airgap, we have come up with an idea of somewhat shorter inserts: in such a way it is possible to effectively manipulate the magnetic flux in the airgap by exposing parts of the permanent magnets at the ends. This leads to higher $B_z$ at the ends and thus can compensate the decrease in field strength off the airgap center. Figure 1d shows experimental curves of the $B_z$ change ($\Delta B_z$) on the Y axis in the airgap center ($X, Z = 0$) for different lengths of the steel inserts. These data demonstrate over 2 orders of magnitude improvement in the field homogeneity in the central part (50% of $L_{mag} = 40$ mm), which was obtained for the optimal length of the steel inserts (33.6 mm, 84% of $L_{mag}$). Inset in Fig. 1d shows how the field uniformity can be tuned after magnet assembly by displacing the steel inserts. Interestingly, by such fine tuning, it is possible to achieve sufficient field homogeneity in a region somewhat off the magnet center even for not optimal insert length: for example, the curve for the -0.05 mm off-center displacement demonstrates $\Delta B_z < 50$ ppm in a rather long interval of 8 mm (20% of $L_{mag}$). Such off-center position is beneficial for flow measurements as the opposite longer part of the magnet can be fully used for pre-magnetizing the inflow liquid [26].

***C-shape magnet.*** – Based on our experimental findings and COMSOL® simulations, we propose an optimized C-shape magnet design depicted in Figure 2a,b. The permanent magnets are SmCo, the soft-magnetic insert plates and the external flux-guide part (yoke) are made of magnetic stainless steel 410. As the magnetic flux from the permanent magnets is mostly enclosed in the yoke, this substantially increases the field inside of the airgap (from around 200 mT to 490 mT) and decreases undesirable stray field outside. Figure 2c shows the field line of the general safety standard of 0.5 mT, which is very close to the magnet and thus can satisfy safety requirements. The "C" shape allows access from the side, and the 5-mm airgap has enough space for additional electronics or microfluidic fixtures inside. By decreasing the airgap height, it is possible to achieve even higher magnetic field, which can be exploited for further field increase and/or miniaturization. Optimal length of the steel inserts leads to a substantial improvement of the longitudinal field homogeneity, as seen in Fig. 2d. These results are fully consistent with the behavior presented in Fig. 1d and discussed therein.

Figure 2d also shows a curve for the configuration without steel inserts in the airgap: the curve has local irregularities, which is a notorious problem usually attributed to the nonuniform magnetization distribution in the permanent magnets [27]. The steel inserts effectively smooth out these local irregularities, though their footprint (much weaker in magnitude) can be still seen in case of our 1-mm steel inserts; cf. insert to Fig. 2d. Figure 2e compares lateral (vs. $X$) and vertical (vs. $Z$) magnetic field variations (in the magnet center, $Y = 0$) for the configurations with and without steel inserts: for the latter case (bottom panel in Fig. 2e), besides much larger field inhomogeneity, the characteristic field maximum/minimum in the $X/Z$-dependences are offset from the initial central position. What is more, if one of the magnets is rotated by 180°, the curve acquires very steep slope (see Fig. 2f) indicating that the magnets have a substantial magnetization gradient in the longitudinal direction. This gradient is compensated in the default configuration. Same behavior was observed for a few different sets of magnets from the same batch; sets 1, 2, 3 in Fig. 2f. The optimized steel inserts have improved the field homogeneity for five made prototypes with different permanent magnet sets to the same extent as presented here, indicating high reproducibility of the results. Therefore, even in case of such imperfect permanent magnets, our design can provide sufficient field homogeneity.

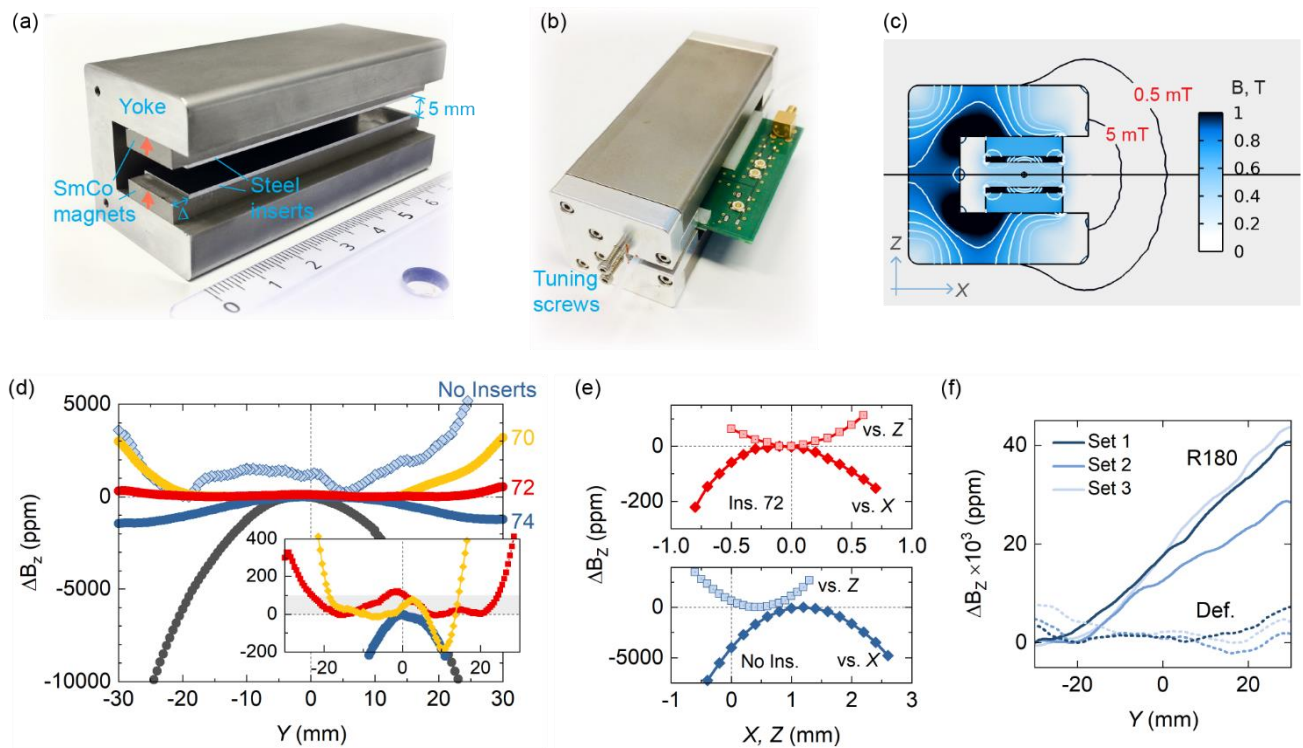

**Fig. 2.** Optimized C-shape permanent magnet. (a) Physical prototype and its main components. (b) Assembled magnet with aluminum fixtures for simple assembly and tuning the longitudinal field homogeneity (by displacing the inserts with side screws; parameter Δ in panel a). (c) $XZ$ cross-section view of the equipotential **B** field contours obtained by the FEM simulations (slice at $Y = 0$, magnet center). (d) Longitudinal field change $\Delta B_Z$ in the center of the airgap ($X$, $Z$ = 0) for different insert lengths $L_{ins}$ = 80, 74, 72, and 70 mm and without inserts. (e) Lateral (vs. $X$) and vertical (vs. $Z$) $\Delta B_Z$ in the magnet center ($Y = 0$) for $L_{ins}$ = 72 mm (top) and without inserts (bottom). (f) Longitudinal $\Delta B_Z$ (no inserts) for different sets of permanent magnets and for the case when one of the magnets is rotated by 180° in $XY$ plane ($Y$ direction is changed to $-Y$).

## 3. NMR experiments

***NMR field mapping.*** – Figure 3a shows an FID (Free Induction Decay) curve obtained for deionized water using our C-shape magnet prototype ($L_{ins}$ = 70 mm) and a solenoid coil probe (sensing volume 1.6 mm i.d., 1.8 mm long) after an optimal 12 μs excitation pulse (P90 pulse in the NMR terminology). The decay time of the FID (conventionally marked as $T_2^* \approx 0.5$ ms) is much shorter than the relaxation time of water at room temperature, $T_2 \approx 2.5$ s. This is a direct consequence of the field inhomogeneity since the excited nuclear spins in different regions of the sensing volume are processing in local magnetic field slightly different in strength: the worser the homogeneity, the faster the spins lose their coherence after excitation. Fast Fourier Transform (FFT) of the FID gives a resonance peak with central resonance frequency $f \approx 20.8$ MHz, which for proton resonance can be converted into $B \approx 490$ mT. The obtained linewidth at half maximum, LWHM ≈ 200 ppm, which is the quantitative characteristic of the field distribution in the sensing volume, is meeting desired specifications even though the used probe had much larger i.d. than the target 1.0 mm.

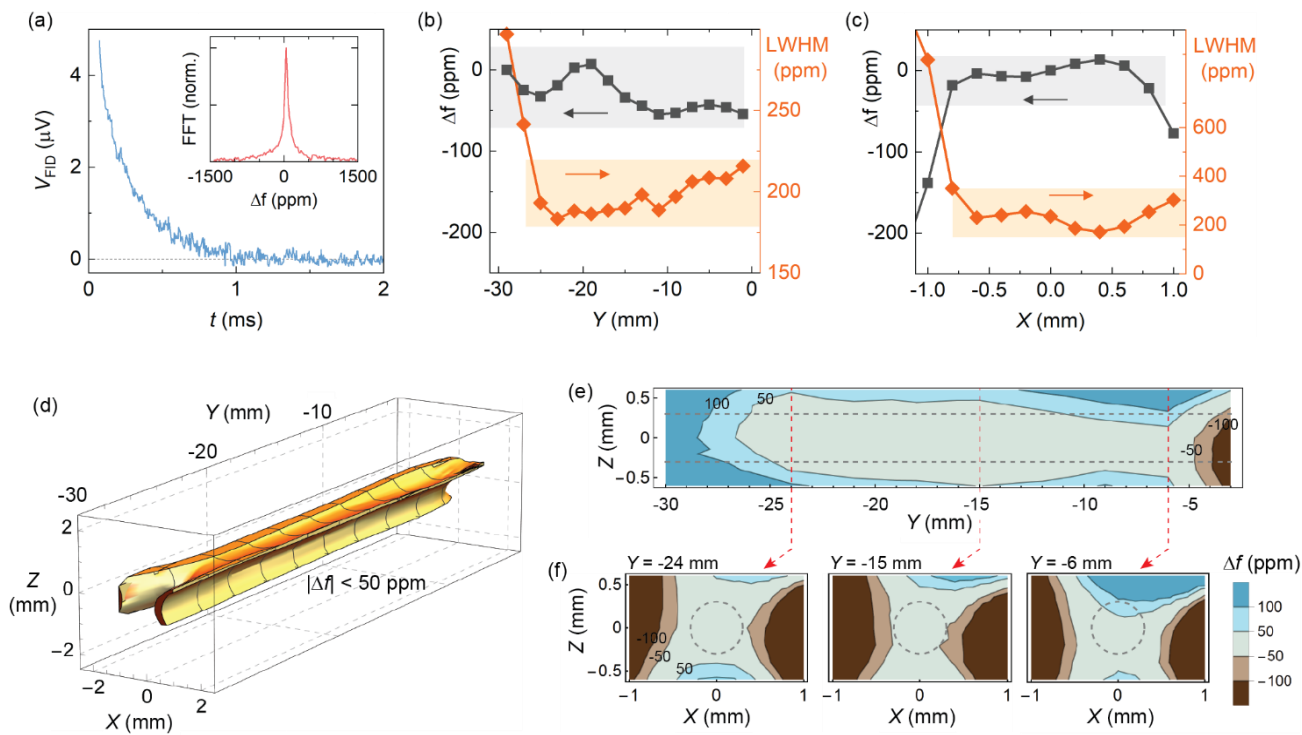

**Fig. 3.** FID measurement and NMR field mapping. (a) FID measured for deionized water using a solenoid coil probe (sensing volume 1.6 mm i.d., 1.8 mm long) and corresponding FFT resonance peak (inset). (b),(c) Relative change in peak central frequency (black squares; left tick axis) and peak LWHM (orange diamonds; right tick axis) versus longitudinal $Y$ position ($X$, $Z$ = 0; panel b) and lateral $X$ position ($Y$, $Z$ = 0; panel c). (d) 3D field map of region of interest with $|\Delta f|$ < 50 ppm and selected 2D slices in planes YZ ($X$ = 0; panel e) and XZ (panel f). The map was obtained using a small coil with sensing volume 0.6 mm i.d. and 0.5 mm long and copper sulfate water solution inside. $X$, $Y$, $Z$ = 0 correspond to the airgap center.

Figure 3b demonstrates how $f$ (black) and LWHM (orange) change when the probe is moved along the magnet airgap (along axis Y; $X$, $Z$ = 0). Figure 3c shows the change in $f$ and LWHM in the lateral $X$-direction at $Y$ = −5 mm. These NMR data remarkably reproduce what was obtained using the Hall probe; cf. Fig. 2c.

Employing even smaller coil probe (sensing volume 0.6 mm i.d., 0.5 mm long) with copper sulfide water solution as the medium (its $T_2$ = 50 ms enables faster measurements), it was possible to obtain a 3D field map in the region of interest. Figure 3d shows the 3D region where $|\Delta f|$ < 50 ppm (or equally $|\Delta B|$ < 50 ppm); Figures 3e,f gives selected 2D slices of this region in YZ (panel e) and XZ (panel f) planes. Such field mapping gives more detailed information on the field distribution in the region of interest and confirms sufficient field homogeneity of our C-shape magnet.

***Test flow measurements.*** – Since our C-shape magnet was designed for micro-flow measurements and its size is defined by the desired sensing volume (1 mm i.d., 20 mm long), we have fabricated an NMR probe having a long solenoid coil with the sensing volume 1.6 mm i.d. by 20 mm long. Figure 4b shows the probe with the coil wound around a 2.0 mm o.d. glass tube and soldered to a PCB with single-ended impedance matching circuit. All the components are mounted on a 3D-printed holder of the sizes matching the airgap so that the probe can be easily slid into and fixed at the right position inside the magnet. Such a probe design allows quick and easy replacement of probes of different coil sizes.

Here we give an example of flow rate measurements based on NMR relaxometry by measuring $T_2$ relaxation curves for various flow rates; more details about the measurement method are in Ref. [28]. The $T_2$ curves undergo changes for different flow rates because of the outflow of the excited spins from the sensing coil volume. The same coil excites the spins by P90 pulse and then recovers the coherence of the spins by a sequence of P180 pulses, while measuring the response in-between; such classic spin-echo signal is depicted in Fig. 4a. In our setup, P180 pulses have the same pulse duration as P90 (40 μs) but twice larger amplitude and twice longer time between the pulses. Because of the choice of such pulse sequence, the first spin-echo peak is lower in amplitude than the second one; all the further peaks gradually decrease in amplitude owing to spin relaxation. Maximums of the spin-echo peaks build a $T_2$ curve, examples of which measured for deionized water and different flow rates are shown in Fig. 4c. The static relaxation curve fits perfectly by exponential decay function exp(-$t/T_2$) with $T_2$ = 2.25 s, which is close to the look-up table value for water at room temperature.

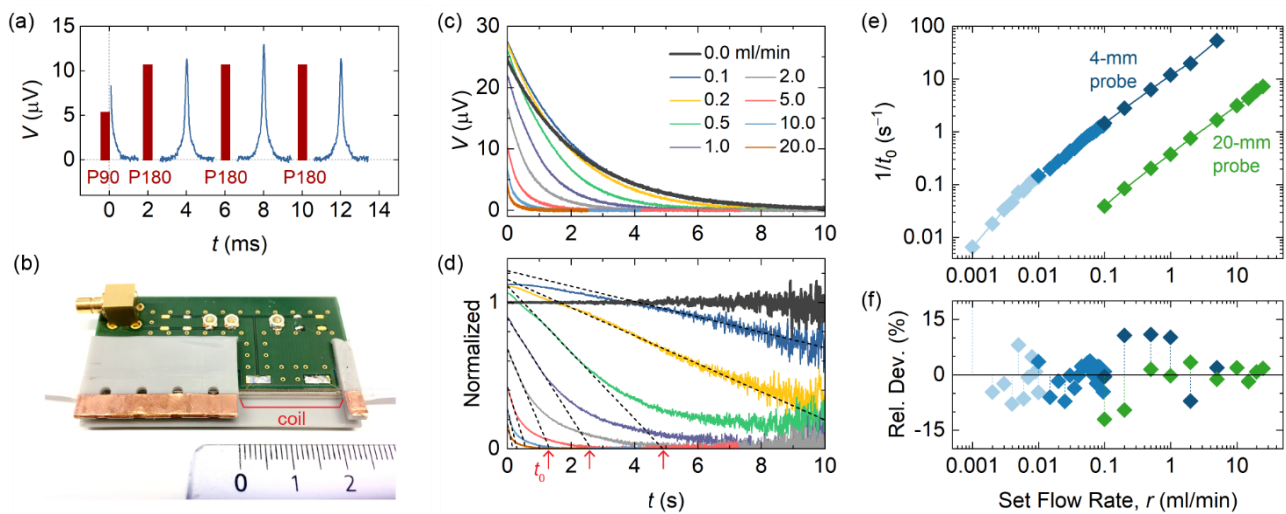

**Fig. 4.** Flow rate measurements of deionized water. (a) Example of measured spin-echo peaks using CPMG rf pulse sequence (schematically depicted as vertical bars) and (b) a 20-mm long solenoid coil secured on the probe holder with single-ended impedance matching PCB circuit. (c) Typical $T_2$ relaxation curves measured for different flow rates and (d) corresponding normalized curves divided by an exponential fit of the static $T_2$ curve. The linear part of each normalized curve is fitted and outflow time $t_0$ is determined as the crossing of the time axis. (e) Obtained $1/t_0$ versus set flow rate, $r$, for the 20-mm (i.d. 1.6 mm) and 4-mm (i.d. 0.6 mm) probes. (f) Corresponding relative deviations from a linear fit. The three data sets for the 4-mm probe correspond to different settings (syringe sizes) of the used syringe pump.

It is important to note that the time between P180 pulses (2 $\tau$) in our measurements was rather short, $\tau$ = 0.1 ms, whereas choosing more typical $\tau$ = 1 ms decreases (effective) $T_2$ to around 1 s (not shown). This is the consequence of a finite field homogeneity and interdiffusion of spins to the regions of slightly different field strength in the sensing volume: the longer the spins are without the external 'stimuli' (P180 pulse), the faster they lose coherence. As a result, the relaxation curve decays faster although the spins are not fully relaxed but decoherent when the curve flattens out (at around 3$T_2$), that is why it is appropriate to use notation *effective* $T_2$. Choosing $\tau$ = 0.1 ms effectively suppresses the effect of spin diffusion on our measurements.

The flow component can be extracted from the measured relaxation curves by dividing them by the static curve (for better SNR, by its exponential fit), as shown in Fig. 4d. Only the first part of the normalized curves (top half) is linear, whereas the second part (bottom half) is tailing off. Such behavior is typical for a laminar flow, which is the case for thin capillaries, and explained by the flow velocity profile: the flow velocity is maximum in the center (2 times of the average flow speed) and is getting slower in radial layers closer to the capillary wall. It means that, when the majority of the excited spins in the central part is out of the sensing volume, the slowly flowing spins at the wall are still contributing to the signal.

To correlate obtained curves with corresponding flow rates, it is helpful to fit the linear part of the curves and find a correlation parameter – e.g. time $t_0$, when the linear fit crosses the time axis, is a convenient parameter to use. Within such approach, from Fig. 4d, it is well seen that the lower flow rate determination is limited by the increased noise at later times (the noise is technically amplified by dividing experimental curves by values close to 0), whereas the upper flow rate is limited by a substantial decrease in the initial amplitude. The latter is owing to the fact that inflowing spins do not have enough time to align along the **B**$_0$ field of the magnet, i.e. the upper flow rate is limited by the premagnetization length of the magnet (in our case, since the probe was shifted to one end, this length is around 40 mm, $0.5L_{mag}$).

Figure 4e shows obtained $1/t_0$ versus set flow rates for two probes of different lengths (4 mm and 20 mm) and i.d. (0.6 mm and 1.6 mm, respectively). The offset between the curves is because the actual linear flow velocities are different for same flow rates; the flow velocity is higher in capillaries with smaller i.d. The larger 20-mm probe allowed measurements of flow rates up to 20 ml/min (flow speed 165 mm/s), whereas the smaller 4-mm probe extended the range down to 0.001 ml/min (0.058 mm/s). Overall, these two probes cover the range over 4 orders of magnitude. These data illustrate how choosing an appropriate probe size makes it possible to work in a desired flow rate range.

The data sets for the 4-mm probe (marked by different color tones in Fig. 4e,f) were obtained using a syringe pump supplied with syringes of different sizes for optimal pump functioning in different flow rate ranges. The relative deviations from linear fits, shown in Fig. 4e, having the average less than 5%, can be further decreased by averaging, temperature stabilization, and improving data processing.

It is noteworthy that presented data were obtained using single-shot measurements in the ambient environment at room temperature without any temperature stabilization or temperature compensation. Temperature variations, and especially temperature gradients over position, can affect the magnetic field strength and its homogeneity, since the temperature coefficient of the used magnetic materials is around 200-300 ppm/K. However, in the current design, large temperature gradients are avoided by the one-piece steel yoke having high thermal conductivity and thus quickly redistributing the heat over the whole magnet. Moreover, our simulations demonstrate that the temperature variations affect the field strength but leave the field homogeneity basically unchanged.

## 4. Discussion and Conclusions

Proposed C-shape magnet design, efficient and simple in assembly, can be a starting point for further improvements, some of which we discuss here on the basis of our simulations. For example, by increasing the height and width of the SmCo magnets and the yoke, resulting in a magnet with total size not larger than 6x6x8 $cm^3$, the field strength of 1 T can be achieved, which e.g. can quadruple the SNR. On the other hand, as seen in Fig. 2b, the magnetic flux is mostly focused in certain regions of the yoke: our numerical studies (not shown) indicate that optimizing the yoke shape can lead to at least 25% additional decrease in size and weight.

The key component behind the much improved field homogeneity in our magnet design is the magnetic steel inserts having a shape of simple rectangular flat plate. By tapering an inner side of these plates in a special manner, it would be possible to come up with even better field homogeneity or, alternatively, with same field homogeneity but much smaller magnets. This approach is somewhat similar to the methods based on shaped pole pieces [29], which can be also 3D printed [30]. However, employing such tapering to construct a magnet with considerably better homogeneity (below 10 ppm) can be problematic owing to the high sensitivity to manufacturing and material imperfections (cf. Fig. 2c-e). Our simulations indicate that 10-μm errors in the shape or position of such tapered inserts lead to noticeable field distortions. On the other hand, our numerical experiments show that a few-ppm field homogeneity, but less sensitive to fabrication and material imperfections, is possible with a bit more complex design based on thick prism-like pole pieces combined with a specific permanent magnet arrangement (work in progress; will be reported elsewhere).

Our test flow measurements demonstrate that the magnet is suitable for sensing flow rates in a wide range, from 0.001 to 20 ml/min. The measurements of high flow rates are technically limited by the premagnetization length. In fact, the premagnetization section can be made separately (having even simpler and more compact design) and when long enough can potentially extend the range up to 100 ml/min. At the same time, the main magnet can be made at least twice shorter (<40 mm) while preserving sufficient field homogeneity (<200 ppm) in the desired sensing volume (roughly 1 mm i.d., 20 mm long; $0.5L_{mag}$). The low range limit is essentially determined by miniaturizing NMR probes. Owing to the design scalability, the magnet can be made even more compact for smaller probes.

In summary, we have proposed and demonstrated a scalable magnet design suitable for micro-fluidic flow measurements based on the NMR relaxometry principle. The sensing volume and broad range of flow rates are in the desired range of micro- and nanofluidic applications. Its compact form factor, simplicity in assembly and tuning are key for developing portable multi-phase NMR-based flow meters for laboratory, field and industrial settings. Furthermore, this approach can be advanced for designing compact permanent magnet systems for low-resolution NMR spectroscopy.

**Declaration of Competing Interest**

The authors declare that they have no known competing financial interests or personal relationships that could have appeared to influence the work reported in this paper.

**Acknowledgements**

The authors thank Jankees Hogendoorn, Lucas Cerioni, Marco Zoeteweij, and Koert Kriger of Krohne New Technologies BV, Eren Aydin of Delft University of Technology, and Rob Dierink of the technical center (TCO) at the University of Twente for their input and support.

**Appendix A. Materials and Methods**

**Magnet materials and assembly.** The assembled magnet prototypes have been built using custom-made permanent magnets $Sm_2Co_{17}$ (Schallenkammer Magnetsysteme BmbH, Germany) and magnetic stainless steel 410 (Salomon's Metalen B.V., the Netherlands): the inserts were laser-cut from a 1-mm thick plate, and the yoke was CNC machined from a 60-mm diameter round bar. The inner bed for the inserts and SmCo magnets as well as the side fixtures were made of aluminum. The prototypes were manufactured at the Technical Center for Education and Research (TCO) of the University of Twente.

**Field mapping.** The magnetic field inside the magnets were measured using a magnetometer (Teslameter 3002, Project Elektronik GmbH, Germany) equipped with a 0.1×0.1-mm$^2$ active area Hall probe (Transverse Probe T3-1,4-5,0-70). The 3D field maps were measured by employing compact motorized translation stages (Thorlabs, Inc.) mounted in 3-axis XYZ configuration. NMR field mapping was carried out using a solenoid coil probe with active cylindric volume 0.6 mm diameter and 0.5 mm long, filled with a copper sulfide water solution (CuSO$_4$ concertation 8 mg/ml; effective $T_2 \approx 50$ ms).

**NMR measurements** were performed by means of commercial electronics (Benchtop MRI unit, Pure Devices GmbH, Germany) and in-house made solenoid coils and impedance matching circuits. The coils of different lengths were wound around glass capillaries using a 32-μm diameter enameled copper wire fixed with a UV light curing glue. The relaxation curves were measured using CPMG pulse sequence with the pulse duration from 10 μm to 40 μm (respectively for the 4-mm and 20-mm long coils) and the echo time between the P180 pulses of 0.2 ms; the sample frequency was 200 kHz. The amplitudes of the P90 and P180 pulses were 1.9 V and 3.8 V, respectively. Different flow rates were supplied using a syringe pump (the Harvard Apparatus PHD 22/2000) and PTEF tubing.

**COMSOL® simulations** were performed using AC/DC module, software version 5.6. The static magnetic field distribution was modeled using Magnetic fields, No Currents interface: Remanent flux density model for the permanent magnets (partially accounts for demagnetization and have best agreement with the experiments) and B-H curve model for the soft-magnetic steel inserts and yoke. It is noteworthy that (fixed) Magnetization model used for the permanent magnets gives similar results, whereas Nonlinear permanent magnet model, often used for magnets to account for (shape) demagnetization, gives rather large discrepancy with the performed experiments.

**References**

[1]    S. Fan, Q. Zhou, K.-M. Lei, P.-I. Mak, R.P. Martins, Miniaturization of a Nuclear Magnetic Resonance System: Architecture and Design Considerations of Transceiver Integrated Circuits, IEEE Transactions on Circuits and Systems I: Regular Papers. 69 (2022) 3049–3060. https://doi.org/10.1109/TCSI.2022.3187041.

[2]    J. Anders, F. Dreyer, D. Krüger, I. Schwartz, M.B. Plenio, F. Jelezko, Progress in miniaturization and low-field nuclear magnetic resonance, Journal of Magnetic Resonance. 322 (2021) 106860. https://doi.org/10.1016/j.jmr.2020.106860.

[3]    S.S. Zalesskiy, E. Danieli, B. Blümich, V.P. Ananikov, Miniaturization of NMR Systems: Desktop Spectrometers, Microcoil Spectroscopy, and "NMR on a Chip" for Chemistry, Biochemistry, and Industry, Chem Rev. 114 (2014) 5641–5694. https://doi.org/10.1021/cr400063g.

[4]    B. Blümich, Low-field and benchtop NMR, Journal of Magnetic Resonance. 306 (2019) 27–35. https://doi.org/10.1016/j.jmr.2019.07.030.

[5]    D. Ha, J. Paulsen, N. Sun, Y.-Q. Song, D. Ham, Scalable NMR spectroscopy with semiconductor chips, Proceedings of the National Academy of Sciences. 111 (2014) 11955–11960. https://doi.org/10.1073/pnas.1402015111.

[6]    K.-M. Lei, H. Heidari, P.-I. Mak, M.-K. Law, F. Maloberti, R.P. Martins, A Handheld High-Sensitivity Micro-NMR CMOS Platform With B-Field Stabilization for Multi-Type Biological/Chemical Assays, IEEE J Solid-State Circuits. 52 (2017) 284–297. https://doi.org/10.1109/JSSC.2016.2591551.

[7]    N. Sahin Solmaz, M. Grisi, A. v. Matheoud, G. Gualco, G. Boero, Single-Chip Dynamic Nuclear Polarization Microsystem, Anal Chem. 92 (2020) 9782–9789. https://doi.org/10.1021/acs.analchem.0c01221.

[8]    D. Sakellariou, G. le Goff, J.-F. Jacquinot, High-resolution, high-sensitivity NMR of nanolitre anisotropic samples by coil spinning, Nature. 447 (2007) 694–697. https://doi.org/10.1038/nature05897.

[9]    R.M. Fratila, A.H. Velders, Small-Volume Nuclear Magnetic Resonance Spectroscopy, Annual Review of Analytical Chemistry. 4 (2011) 227–249. https://doi.org/10.1146/annurev-anchem-061010-114024.

[10]   N. Sun, T.-J. Yoon, H. Lee, W. Andress, R. Weissleder, D. Ham, Palm NMR and 1-Chip NMR, IEEE J Solid-State Circuits. 46 (2011) 342–352. https://doi.org/10.1109/JSSC.2010.2074630.

[11]   H. Ryan, S.-H. Song, A. Zaß, J. Korvink, M. Utz, Contactless NMR Spectroscopy on a Chip, Anal Chem. 84 (2012) 3696–3702. https://doi.org/10.1021/ac300204z.

[12]   H. Lee, E. Sun, D. Ham, R. Weissleder, Chip–NMR biosensor for detection and molecular analysis of cells, Nat Med. 14 (2008) 869–874. https://doi.org/10.1038/nm.1711.

[13]   J. Mitchell, L.F. Gladden, T.C. Chandrasekera, E.J. Fordham, Low-field permanent magnets for industrial process and quality control, Prog Nucl Magn Reson Spectrosc. 76 (2014) 1–60. https://doi.org/10.1016/j.pnmrs.2013.09.001.

[14]   Bruker's Ascend 1.2 GHz/28.2 T NMR instrument, (n.d.). https://www.bruker.com/en/products-and-solutions/mr/nmr/ascend-ghz-class.html (accessed October 19, 2022).

[15]   A. McDowell, E. Fukushima, Ultracompact NMR: 1H Spectroscopy in a Subkilogram Magnet, Appl Magn Reson. 35 (2008) 185–195. https://doi.org/10.1007/s00723-008-0151-3.

[16] E. Danieli, J. Perlo, B. Blümich, F. Casanova, Small Magnets for Portable NMR Spectrometers, Angewandte Chemie International Edition. 49 (2010) 4133–4135. https://doi.org/10.1002/anie.201000221.

[17] A. Bogaychuk, V. Kuzmin, Accounting for material imperfections in the design and optimization of low cost Halbach magnets, Review of Scientific Instruments. 91 (2020) 103904. https://doi.org/10.1063/5.0013274.

[18] A.J. Parker, W. Zia, C.W.G. Rehorn, B. Blümich, Shimming Halbach magnets utilizing genetic algorithms to profit from material imperfections, Journal of Magnetic Resonance. 265 (2016) 83–89. https://doi.org/10.1016/j.jmr.2016.01.014.

[19] H.Y. Carr, E.M. Purcell, Effects of Diffusion on Free Precession in Nuclear Magnetic Resonance Experiments, Physical Review. 94 (1954) 630–638. https://doi.org/10.1103/PhysRev.94.630.

[20] S. Meiboom, D. Gill, Modified Spin-Echo Method for Measuring Nuclear Relaxation Times, Review of Scientific Instruments. 29 (1958) 688–691. https://doi.org/10.1063/1.1716296.

[21] R. Skomski, J.M.D. Coey, Magnetic anisotropy — How much is enough for a permanent magnet?, Scr Mater. 112 (2016) 3–8. https://doi.org/10.1016/j.scriptamat.2015.09.021.

[22] K. Chonlathep, T. Sakamoto, K. Sugahara, Y. Kondo, A simple and low-cost permanent magnet system for NMR, Journal of Magnetic Resonance. 275 (2017) 114–119. https://doi.org/10.1016/j.jmr.2016.12.010.

[23] E. Potenziani, H. Leupold, Permanent magnets for magnetic resonance imaging, IEEE Trans Magn. 22 (1986) 1078–1080. https://doi.org/10.1109/TMAG.1986.1064465.

[24] COMSOL Multiphysics® v. 5.6 www.comsol.com. COMSOL AB, Stockholm, Sweden, (n.d.).

[25] AC/DC Module User's Guide, pp. 238-362. COMSOL Multiphysics® v. 5.6. COMSOL AB, Stockholm, Sweden. 2020, (n.d.).

[26] In NMR of liquids, the nuclear spins need a rather long characteristic time to align along the applied magnetic field – spin-lattice relaxation time, $T_1$, which for example can last for a few seconds for tap water at room temperature. Ideally, the nuclear spins should be well magnetized when entering the NMR probe placed in the magnet center.

[27] L.A. and J.G.E.G. Y. P. Klein, Influence of the Distribution of the Properties of Permanent Magnets on the Field Homogeneity of Magnet Assemblies for Mobile NMR, IEEE Trans Magn. 57 (2021) 1–7.

[28] E. Aydin, K.A.A. Makinwa, A Low-Field Portable Nuclear Magnetic Resonance (NMR) Microfluidic Flowmeter, in: 2021 21st International Conference on Solid-State Sensors, Actuators and Microsystems (Transducers), IEEE, 2021: pp. 1020–1023. https://doi.org/10.1109/Transducers50396.2021.9495479.

[29] A.E. Marble, I. v. Mastikhin, B.G. Colpitts, B.J. Balcom, A constant gradient unilateral magnet for near-surface MRI profiling, Journal of Magnetic Resonance. 183 (2006) 228–234. https://doi.org/10.1016/j.jmr.2006.08.013.

[30] B.M.K. Alnajjar, A. Buchau, L. Baumgärtner, J. Anders, NMR magnets for portable applications using 3D printed materials, Journal of Magnetic Resonance. 326 (2021) 106934. https://doi.org/10.1016/j.jmr.2021.106934.